\documentclass[twoside]{article}
\usepackage{fleqn,espcrc2}
\usepackage{epsfig}

\title{ Electronic structure and magnetic anisotropy of the  
[Co$_4$(hmp)$_4$(CH$_3$OH)$_4$Cl$_4$] molecule }
\author{ Tunna Baruah 
\address{Department of Physics, Georgetown University, Washington, DC 20057, USA }
and Mark R. Pederson \thanks{Corresponding author: FAX: +1-202-404-7546;
           e-mail: pederson@dave.nrl.navy.mil} 
\address{Center for Computational Materials Science,
Code 6392,
 Naval Research Laboratory 
 Washington, DC 20375-5000, USA }
}
\begin{document}

\begin{abstract}

Accurate density-functional based calculations have been performed on 
the
Co$_4$(hmp)$_4$(CH$_3$OH)$_4$Cl$_4$ molecular magnet where hmp is
deprotonated hydroxymethyl pyridine. In addition to the 
experimentally observed staggered geometry, we identify two isomers, 
referred to as eclipsed and half-staggered/half-eclipsed, that are 
reasonably low in energy. Our calculations show that the magnetic 
anisotropy is strongly dependent on the pyridine-pyridine separation
and that the three structures exhibit easy axis, easy plane and triaxial 
behavior. Other effects such as partial reprotonation of the hmp
is considered.

\end{abstract}

\maketitle

The magnetic molecules containing transition metal atoms are being 
widely studied  due to potential technological applications 
for information  storage and quantum computing \cite{Leuenberger}. 
The magnetic 
molecules show magnetic hysteresis behavior reminiscent of single 
domain magnets and exhibit the phenomenon of quantum tunneling of 
magnetization.  Observation of such behavior in a molecular magnet 
is greatly facilitated by a reasonably high net spin and a large 
magnetic anisotropy  energy (MAE). A very recent experimental report 
of a single molecule magnet consisting of 
Co$_4$(hmp)$_4$(CH$_3$OH)$_4$Cl$_4$ where 
hmp- is the deprotonated hydroxymethylpyridine, suggests that  this 
molecule is quite promising since the magnetic anistropy energy per 
transition metal atom is high  (~25-50K) \cite{Yang} compared to other magnetic 
molecules Fe$_8$-tacn \cite{Fe1}, Mn$_{12}$-acetate \cite{Mn1,Mn2,Mn3,Mn4} where it is  
(~3-6K). The reported 
ferromagnetic ordering of the Co$_4$ molecule also differs qualitatively 
from the ferrimagnetic spin ordering observed in the Mn$_{12}$ and Fe$_8$. 
The negative anisotropy energy for a transition metal with more than 
5 d electrons is also considered to be unusual.

    In order to gain more insight into the properties of this molecule,
 we have carried out a detailed  {\it ab initio}  study of the electronic 
structure and magnetic anisotropy energy of 
[Co$_4$(hmp)$_4$(CH$_3$OH)$_4$Cl$_4$] 
for different conformers. Density functional theory \cite{DFT1,DFT2} based  
all-electron,  spin-polarized calculations were carried out with the 
NRLMOL code \cite{NRLMOL1,NRLMOL2} within the generalized gradient 
approximation to the 
exchange-correlation functional \cite{PBE}. 
%The MAE was calculated according to 
%the methods  given in Ref. \cite{Ped-Khanna}  and Ref. \cite{Kor-ped} .

\begin{figure}[h]
\epsfig{file=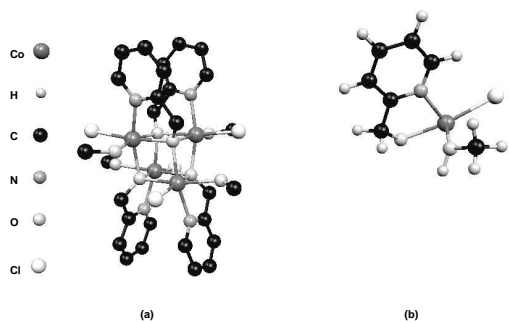 ,width=1.0\linewidth,clip=true}
\caption
{ (a) The  geometry of the lowest-energy staggered structure
in which the hydrogen atoms are not shown. (b) The inequivalent building
block of each molecule which 
consists of the Co$^{2+}$, Cl$^-$, CH$_3$OH and the hmp$^-$ ligands.}
\label{geom}
\end{figure}

\begin{figure}[h]
\epsfig{file=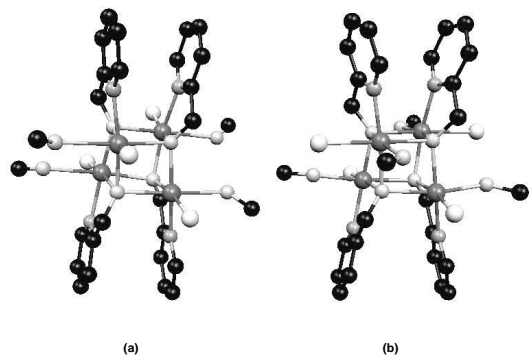 ,width=1.0\linewidth,clip=true}
\caption
{  The optimized  (a) eclipsed and (b) half-staggered 
half-eclipsed geometries in which the hydrogen atoms are not shown.}
\label{geom2}
\end{figure}

   In the experimentally obtained Co$_4$ cluster \cite{Yang}, the four 
cobalt atoms in the molecule are bonded to organic 
hydroxy-methyl-pyridine(hmp) ligands (Fig. \ref{geom}(a)).  The four 
Co(II) atoms and the four oxygens from the CH$_2$O$^-$ radicals of 
the hmp ligand  form an inner cubane structure.  The 
experimentally observed molecule exhibits S$_4$ point group 
symmetry and is made up of four  units where each 22-atom 
unit (Fig. \ref{geom}(b)) consists of Co$^{+2}$, Cl$^-$, CH$_3$OH and the hmp$^-$ 
ligand.  An eclipsed geometrical isomer (Fig. \ref{geom2}(a)) can be generated by 
continuously distorting the lower half of the staggered 
experimental geometry or  from the inequivalent 22-atom 
complex from the group of order 4 generated from (xyz) $\longrightarrow$ 
(-x,-y,z) and (xyz) $\longrightarrow$ (y,x,-z).  Another energetically 
competive structure (Fig. \ref{geom2}(b)), which we refer to as half staggered 
and half eclipsed (HSHE), can be generated by rotating and 
appropriately reorienting only the Cl and ligands on the lower 
half of the experimental structure. This structure has one 
symmetry operation [(xyz)$\longrightarrow$  (x,-y,-z)]. Complete geometry 
optimizations were carried out for each conformer. 

    The starting parameters for the experimentally reported 
staggered  geometry were taken from the reported  isostructure 
of the [Ni$_4$(hmp)$_4$(CH$_3$OH)$_4$Cl$_4$] \cite{polyhedron} 
molecular crystal. Upon 
relaxation, the volume of the molecule increases due to the 
repulsion of the hmp ligands in parallel positions.  The 
eclipsed structure was found to be higher in energy than the 
staggered structure  by 2.77  eV.  Since this energy difference 
is probably due to intraplanar and interplanar steric repulsions, 
we have been motivated to look at the half-staggered and half-eclipsed geometry 
as well.  We find that the relaxed HSHE geometry is only 0.79 eV above the 
experimental geometry. The cohesive energy was found to be approximately 
4.7 eV/atom for all structures.  

\begin{figure}[h]
\epsfig{file=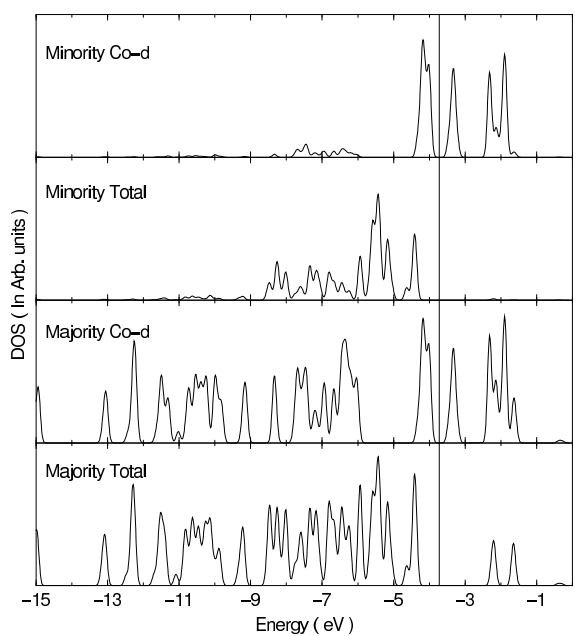 ,width=1.0\linewidth,clip=true}
\caption
{  The total  and Co-d majority and minoroty spin density of states of the
staggered Co$_4$ molecule.   }
\label{DOS}
\end{figure}

 The spin-projected total density of states (DOS) as well as
the Co d density of states  of the staggered structure 
are shown in Fig.  \ref{DOS}. 
The Fig. \ref{DOS} clearly shows that the states near the
Fermi level arises mainly from the Co atoms. These states also contain 
 minor contributions
from the oxygen and chlorine atoms. The Co d majority states are fully occupied.
The figure  shows that both the HOMO
and the LUMO belongs to the minority spin Co {\it d} levels which is true
in all
the structures. 
 The minority gap in the staggered structure is 0.55 eV.
A smaller gap
leads to a higher value of the magnetic anisotropy energy. The gap
is found to be highly  sensitive to the orientation of the ligands.

     The spin-density in a sphere of radius 2.6 a.u. around each 
Co atom shows that Co atoms have a local moment of 3 $\mu_B$ which 
is also expected from electron counting considerations. This leads to 
a ferromagnetic structure with a total moment of 12 $\mu_B$ (or S=6) for all
 isomers and is in agreement with experiment \cite{Yang}. Therefore, the orientation 
of the hmp ligands does not affect the magnetic moment. An isolated 
Co$_4$O$_4$ cubane is also found to possess a magnetic moment of 12. Our 
investigation on the  effect of hydrogen dopants shows that the 
inclusion of additional hydrogens leads to a partial reprotonation of the 
cubane OC-CH$_2$
groups and insertion of the electrons into the vacant minority Co$_{3d}$ 
states just above the Fermi level. This reduces the magnetic moment by 
1 $\mu_B$  per hydrogen atom.
    
Our calculations have shown that the states near the Fermi  level are 
dominated by the cobalt d states with both the highest occupied and lowest 
unoccupied molecular orbitals (HOMO and LUMO) belonging to the 
minority spin Co d levels.  The HOMO-LUMO gap in the staggered structure 
is 0.55 eV.  The gap is found to be highly  sensitive to the orientation of the 
ligands.  In the  eclipsed structure,  the HOMO-LUMO gap is small (0.09 eV), while 
for the HSHE structure, it is 0.41 eV. This leads to interesting magnetic behavior 
in the three structures. The values of the gap and the  MAE of the three conformers 
are summarized in Table \ref{table1}.

The lowest order spin hamiltonian for a single molecular
magnet can be modelled as
$$ H = D S_z^2 + E(S_x^2 - S_y^2) + g \mu {\vec B}.{\vec S} $$
where D and E are the second-order axial and transverse anisotropies.
${\vec B}$  and g are the applied magnetic field and the Land{\'e}
g-factor, respectively.
The main contribution to the second-order
anisotropy energy arises from the spin-orbit coupling.
The effect of the spin-orbit coupling and the second-order anisotropy
energy was calculated here using the second-order pertubative
treatment presented in Ref. \cite{Ped-Khanna}.  We also used an
alternative treatment given in Ref. \cite{Kor-ped} where
the principal axes  and anisotropy energies
may be determined by diagonalizing the anisotropy tensor.
While the experimental staggered structure with
S$_4$ symmetry is uniaxial,(i.e. E=0 by symmetry) the other
two structures will have a transverse component (E).

   The calculated magnetic anisotropy energy of the uniaxial, optimized 
staggered structure was found to be  23 K although the initial structure, 
based on experimental parameters, showed a high value of MAE of 57 K. The 
separation between pyridine rings was found to increase upon optimization   
which in turn increased the HOMO-LUMO gap.  The HOMO-LUMO gap for the 
initial geometry of the experimental structure was found to be 0.09 eV.  
This observation opens up an exciting possibility of altering the MAE by 
manipulating the pyridine-pyridine separations, in particular, by the 
application of pressure.  It may be mentioned here that Yang  et al. \cite{Yang} 
have established 
that the magnetization barrier is $\sim$ 100-200 K.  The large discrepancy 
between the theoretical and experimental values further supports the 
possibility that the MAE  may exhibit significant pressure dependencies.  
So crystal packing effects may play an important role here.
   
The eclipsed structure, on the other hand,  shows a  more promising 
magnetic behavior (Table \ref{table1}).  In this case, the magnetization has  easy, 
medium and hard axes. The easy axis lies along the axis of the molecule 
as in the experimental structure.  The energy barrier between the easy 
and the medium axes is 95K while that between the medium and hard axes 
is  65K. This leads to a rather large D parameter of -3.55K and an E 
parameter of -0.91K.  Interestingly, the MAE of the HSHE structure  (50K) 
lies   between the MAE of the staggered and the eclipsed structures.  
However,  this geometry has an easy plane of magnetization.  Thus, 
changing the orientation of the ligands significantly, albeit indirectly, 
influences the magnetic behavior of the Co$_4$ through band gap changes.
   
\begin{figure}[h]
\epsfig{file=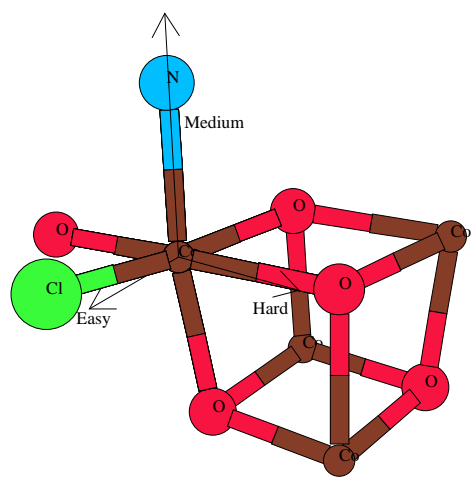 ,width=1.0\linewidth,clip=true}
\caption
{  The local easy, medium and hard axes on a representative Co atom.
Only the cobalt-oxygen cube part of the whole molecule is shown.}
\label{toc}
\end{figure}

One possible way of achieving an overall easy axis is through the orthogonal 
hard axis alignment model in which each cobalt exhibits a local easy plane 
and the intersection of easy planes between the lower and upper Co atoms 
determines an easy axis \cite{Yang}. Another way is for each atom to exhibit a local 
easy axis along the global uniaxis. Within the second-order perturbation 
theory, and the LCAO method used here, it is possible to decompose the 
anisotropy hamiltonian into a sum of 4-center terms by expanding the 
relevant matrix elements into their atomic constituents. Because the Co  d 
states are localized and dominate the behavior near the Fermi level, one 
expects that the single-center diagonal terms will be primarily responsible 
for the anisotropy energy. We have determined numerically that this is an 
excellent approximation and find that in all structures, the local alignments
 are identical with the local hard axis lying along the in-plane Co-O bond 
in the cubane (Co and O are attached to two parallel hmp ligands) and the 
local medium axis lying along the Co-N bond (Fig. \ref{toc}). This leads to the situation 
where the local hard and easy  axes lie on a plane bisecting the molecule. 
For the case of the staggered structure, the local hard axes of the upper 
and lower Co atoms are orthogonal, which leads to the global easy axis lying 
along the axis of the molecule. In the other two structures, the local hard 
axes are aligned parallel and the competition between the medium and the 
local easy axes leads to a triaxial alignment in one and  to an easy plane 
of alignment in the other. For the latter, the easy and the medium axes are 
nearly degenerate.

    In conclusion, we show that the magnetization of the  Co$_4$ molecule 
varies strongly with the orientation of the various ligands.  While the 
staggered and eclipsed structures  have a preferred axis of magnetization, 
the intermediate structure has a preferred plane of magnetization. For the 
higher energy isomer, the energy barrier between the hard and the medium 
axis is quite high ($\sim$ 95K) . This is especially interesting in view of the 
small number of transition metal atoms in  the molecule.   Our results 
suggest that the anisotropy energy may be strongly varied by manipulating 
the pyridine-pyridine spacing and ligand orientation. Determining a means 
for constraining the ligands by either chemical, physical or electrostatic 
means would be a worthwhile investigation.

\section{Acknowledgment}
  The authors gratefully acknowledge the financial support by ONR grant  
N0001400WX2011.

%References ...

%\vspace{3.5in}

%Figures ...

% Tables ...

\begin{table}
\caption{ The magnetic moment($\mu$), HOMO-LUMO gap ($\Delta$), magnetic alignment 
energies, 
with respect to the easy axis,
and type of magnetic alignment of the three different structures of the
Co$_4$ molecule. H and  M  correspond to hard and medium axes respectively. 
S, E and HSHE pertain to staggered, eclipsed and half-staggered/half-eclipsed 
structures respectively. }
\begin{tabular}[t]{lcccccc} \hline
Struc. & $\mu$ & $\Delta$ & \multicolumn{2}{c}{ Energy (K)} & Type  \\   
& &  (eV) &  H &M &  \\   \hline

S & 12 & 0.55 & 23 &23  & Easy  axis    \\ \\
E & 12 & 0.09 & 160 &95  & Triaxial  \\ \\
HSHE & 12  & 0.41 & 50 &10  & $\sim$ Easy  plane \\ \hline

\end{tabular}
\label{table1}
\end{table}

\end{document}